\documentclass{aa}
\usepackage{graphicx,color}
\usepackage{natbib}
\usepackage[english]{babel}
\bibpunct{(}{)}{;}{a}{}{,}

\usepackage{txfonts}
\begin{document}
\newcommand{\re}{\textcolor[rgb]{1,0,0}}

   \title{Circumstellar molecular composition of the oxygen-rich AGB star \object{IK Tau}: I.\ Observations and LTE chemical abundance analysis}


\author   {H. Kim \inst{1,2}  \and F. Wyrowski \inst{1}
           \and K. M. Menten\inst{1} \and L. Decin\inst{3,4}}


   \offprints{H. Kim, e-mail: hyunjoo.kim@aei.mpg.de}

 \institute{
  Max-Planck-Institut f\"ur Radioastronomie,
 		 		        Auf dem H\"ugel 69, 53121 Bonn, Germany
  \and       Max-Planck-Institut f\"ur Gravitationsphysik,
 		 		        Callinstr. 38, 30167 Hannover, Germany
 \and          Instituut voor Sterrenkunde, KU Leuven, Celestijnenlaan 200D, 3001 Leuven,
 Belgium
  \and Sterrenkundig Instituut Anton Pannekoek, University of
  Amsterdam, P.O.\ Box 9429, 1090 CE Amsterdam, The Netherlands}

   \date{received date; accepted date}


  \abstract
   {Molecular lines in the (sub)millimeter wavelength range
     can provide important information about the physical and chemical conditions in the circumstellar envelopes
     around Asymptotic Giant Branch stars.}
   {The aim of this paper is to study the molecular composition in the circumstellar envelope around
   the oxygen-rich star \object{IK~Tau}.}
   {We observed IK Tau in several (sub)millimeter bands using the APEX telescope during three observing periods.
    To determine the spatial distribution of the $\mathrm{^{12}CO(3-2)}$ emission,
    mapping observations were performed. To constrain the physical conditions in the circumstellar envelope, multiple rotational CO emission lines were modeled using
    a non local thermodynamic equilibrium radiative transfer code.
    The rotational temperatures and the abundances of the other molecules
    were obtained assuming local thermodynamic equilibrium.}
   {An oxygen-rich Asymptotic Giant Branch star has been surveyed in the submillimeter wavelength range. Thirty four transitions of twelve molecular species, including maser lines, were detected. The kinetic temperature of the envelope was determined and the molecular abundance fractions of the molecules were estimated.
    The deduced molecular abundances were compared with observations and modeling from the literature and agree within a factor of 10, except for SO$_2$, which is found to be almost a factor 100 stronger than predicted by chemical models.}
   {From this study, we found that IK Tau is a good laboratory to study the conditions in circumstellar
    envelopes around oxygen-rich stars with (sub)millimeter-wavelength molecular lines. 
   We could also expect from this study that the molecules in the circumstellar envelope can be explained more faithful by non-LTE analysis with lower and higher transition lines than by simple LTE analysis with only lower transition lines. In particular, the observed CO line profiles could be well reproduced
by a simple expanding envelope model with a power law structure.}

\authorrunning{H.\ Kim et al.}
\titlerunning{The circumstellar chemistry in the O-rich AGB IK~Tau}

   \maketitle
%

\section{Introduction} \label{introduction}

Stars with initial masses lower than $\sim$8  $\mathrm{M_{\odot}}$
evolve to a pulsationally unstable red
giant star on the Asymptotic Giant Branch (AGB). At this stage, mass
loss from the evolved central star produces an expanding
envelope. Further on, carbon, C, is fused in the core and then oxygen,
O \citep{Yamamura1996,Fukasaku1994}.

AGB stars are characterized by low surface temperatures, $T_{*}\mathrm{\leq
3000}$ K, high
luminosities up to several $\mathrm{10^{4}}$ $\mathrm{L_{\odot}}$, 
and a very large geometrical size up to several AU \citep{Habing1996}. In general, these highly
evolved stars are surrounded by envelopes with expansion velocities between 5
$\mathrm{km\,s^{-1}}$ and 40 $\mathrm{km\,s^{-1}}$. They have high mass-loss
rates between $\mathrm{10^{-8}}$ and $\mathrm{10^{-4}\,M_{\odot} \,yr^{-1}}$.
Their atmospheres provide favorable thermodynamic conditions
for the formation of simple molecules, due to the low temperatures and,
simultaneously, high densities.
Due to pulsation, molecules may reach a distance at which the temperature is
lower than the condensation temperature and at which the density is still high
enough for dust grains to form. Radiation pressure drives the dust away from
the star. Molecules surviving dust formation are accelerated  due to
dust-grain collisions \citep{Goldreich1976}.

The chemistry of the atmospheres and, further out, of the circumstellar envelopes (CSEs) around AGB
stars is dependent on the chemical class. They are classified either as M stars (C/O abundance ratio $<$ 1), S stars (C/O $\approx$
1) or C stars (C/O $>$ 1). The optical and infrared spectra of AGB stars show
absorption from the stellar atmosphere. M-type stellar spectra are
dominated by lines of oxygen-bearing molecules, e.g., the metal oxides SiO and
TiO, and $\mathrm{H_{2}O}$. In C-star atmospheres carbon-bearing molecules
like, a.o., CH, C$_2$, $\mathrm{C_{2}H_{2}}$ and HCN 
are detected at optical and infrared wavelenths, and in the microwave regime \citep[e.g.][]{Loidl2004}.
While the \emph{atmospheric} abundance fractions are nowadays quite well understood in terms of 
initial chemical composition, which may be altered by nucleosynthetic products which are brought to the surface due to dredge-ups,
the main processes determining the \emph{circumstellar} chemical abundance stratification of 
many molecules are still largely not understood. In the stellar photosphere, the high gas density 
ensures thermal equilibrium (TE). Pulsation-driven shocks in the inner wind region suppress TE. 
This region of strong shock activity is also the locus of grain formation, resulting in the depletion 
of few molecules as SiO and SiS. Other molecules, as CO and CS, are thought to be inreactive in the dust 
forming region \citep{Duari1999}. At larger radii, the so-called outer envelope is penetrated by ultraviolet 
interstellar photons and cosmic rays resulting in a chemistry governed by photochemical and ion-molecule reactions. 
This picture on the chemical processes altering the  abundance stratification is generally accepted,
 but many details on chemical reactions rates, molecular left-overs after the dust formation, 
shock strengths inducing a fast chemistry zone etc. are not yet known. 

Spectroscopical studies of molecular lines in the (sub)millimeter range are a very useful tool for estimating
 the physical and chemical conditions in CSEs. Due to its proximity, the carbon-rich AGB star \object{IRC+10216} has 
attracted lot of attention, resulting in the detection of more than 60 different chemical compounds in its CSE \citep[e.g.][]{Ridgway1976Natur.264..345R, Cernicharo2000A&AS..142..181C}. 
Until now, detailed studies of oxygen-rich envelopes have been rare. Recently, \cite{Ziurys2007} have focused on 
the chemical analysis of the oxygen-rich peculiar red supergiant \object{VY~CMa}. VY CMa is, however, not a proto-type of an evolved oxygen-rich star.
 A complex geometry is deduced from Hubble Space Telescope images \citep{Smith2001} 
with a luminosity larger than $\,10^5$\,L$_{\odot}$ and a mass-loss rate of $\sim$$2 \times 10^{-4}$\,M$_{\odot}$/yr 
 \citep{Bowers1983, Sopka1985}.
 VY~CMa is a spectacular object, which because of its extreme evolutionary state can explode as a supernova at any time. 
Interpreting the molecular emission profiles of VY~CMa is therefore a very complex task, subject to many uncertainties. 
To enlarge our insight in the chemical structure in the envelopes of oxygen-rich low and intermediate mass stars,
 we therefore have started a submillimeter survey on the oxygen-rich AGB star \object{IK~Tau}, which is thought to be (roughly) 
spherically symmetric \citep{Lane1987,Marvel2005}. We thereby will advance the understanding on the final stages of stellar 
evolution of the majority of stars in galaxies as our Milky Way and their resultant impact on the interstellar medium and the cosmic cycle.

\subsection{IK Tau}

The Mira variable IK Tau, also known as NML Tau, is located at
$\alpha_{2000}$=$3^{\mathrm{h}}53^{\mathrm{m}}28^{\mathrm{s}}$.8,
$\delta_{2000}$=$11^{\circ}24^{\prime}23^{\prime\prime}$. It was found to be an
extremely cool star having large infrared ($J-K$) excess \citep{Alcolea1999}
consistent with a 2000\,K blackbody. IK Tau shows regular optical variations
with an amplitude of $\sim$ 4.5 mag.

IK Tau is an O-rich star of spectral type ranging from M8.1 to M11.2
\citep{Wing1973}. Its distance was derived by \citet{Olofsson1998} to be 250 pc
assuming a stellar temperature of 2000\,K. The pulsation period is $\sim$470 days \citep{Hale1997}. The
systemic velocity of the star is 33.7 $\mathrm{km\,s^{-1}}$. Mass-loss rate
estimates range from $\mathrm{2.4\times10^{-6}}$ $\mathrm{M_{\odot}\,yr^{-1}}$
\citep[from the CO(J=1-0) line; ][]{Olofsson1998} to $\mathrm{3\times10^{-5}}$
$\mathrm{M_{\odot}\,yr^{-1}}$ \citep[from an analysis of multiple SiO lines;
][]{Gonzalez2003}.

In the circumstellar envelope of IK Tau maser emission from OH
\citep{Bowers1989},
$\mathrm{H_{2}O}$
\citep{Lane1987},
and SiO
\citep{Boboltz2005} and thermal emission
of SiO, CO, SiS, SO, $\mathrm{SO_{2}}$ and HCN have previously been found
\citep{Lindqvist1988,Bujarrabal1994,Omont1993}.
 Obviously, IK Tau is a prime candidate for circumstellar chemistry studies.


\section{Observations}


\begin{table}
\caption{Overview of the molecular line transitions observed with APEX.} 
\label{table:1}      
\centering                          
\begin{tabular}{c c c c}        
\hline\hline                 
   Species   & Transition  & $\nu$ (MHz)  & HPBW (\arcsec) \\      
\hline
   $^{12}$CO &   3 - 2        & 345796.00  &  18        \\
             &   4 - 3        & 461040.78  &  14        \\
             &   7 - 6        & 806651.81  &   8        \\
\hline
   $^{13}$CO & 3$_3$ - 2$_2$  & 330587.94  &  19        \\
\hline
    SiS      &  16 - 15       & 290380.31  &  21        \\
             &  17 - 16       & 308515.63  &  20        \\
             &  19 - 18       & 344778.78  &  18        \\
             &  20 - 19       & 362906.34  &  18        \\
\hline
\hline
$^{28}$SiO      &  7 - 6         & 303926.81  &  20        \\
             &  8 - 7         & 347330.59  &  18        \\
\hline
 $^{29}$SiO  & 7 - 6         & 300120.47   & 20         \\
             & 8 - 7         & 342980.84   & 18         \\
\hline
 $^{30}$SiO  & 7 - 6         & 296575.75   & 21         \\
             & 8 - 7         & 338930.03   & 18         \\
\hline
    SO       & 7$_7$ - 6$_6$  & 301286.13  &  20        \\
             & 8$_8$ - 7$_7$  & 344310.63  &  18        \\
\hline
    SO$_2$   & 3$_{3\, 1}$ - 2$_{2\, 0}$   & 313279.72  & 20   \\
             & 17$_{1\,17}$ - 16$_{0 \,16}$ & 313660.84  & 20   \\
             & 4$_{3\,1}$ - 3$_{2\,2}$     & 332505.25  & 19   \\
             & 13$_{2\,12}$ -12$_{1\,11}$  & 345338.53  & 18   \\
             & 5$_{3\,3}$ - 4$_{2\,2}$     & 351257.22  & 18   \\
             & 14$_{4\,10}$ - 14$_{3\,11}$ & 351873.88  & 18   \\
\hline
   CS        & 6 - 5         & 293912.25   & 21         \\
             & 7 - 6         & 342883.00   & 18         \\
\hline
    HCN      & 4 - 3         & 354505.47   & 18         \\
\hline
   CN        & N=3 - 2, J=5/2 -3/2  & 340031.56  & 18   \\
             & N=3 - 2, J=7/2 -5/2  & 340247.78  & 18   \\
\hline \hline
\multicolumn{4}{c}{\emph{masers}}\\
\hline
  H$_2$O     & 10$_{2\,9}$ - 9$_{3\,6}$    & 321225.63  &  19  \\
             & 5$_{1\,5}$ - 4$_{2\,2}$     & 325152.91  &  19  \\
\hline
 $^{28}$SiO  & v=1, 7 - 6    & 301814.30   &  20  \\
             & v=1, 8 - 7    & 344916.35   &  18  \\
             & v=3, 7 - 6    & 297595.41   &  20  \\
\hline
 $^{29}$SiO  & v=1, 7 - 6    & 298047.33   &  20  \\
\hline
 $^{30}$SiO  & v=1, 8 - 7    & 336602.44   &  19  \\
\hline                                   
\end{tabular}
\tablefoot{The third column lists the transition frequency, the last column the beam size (HPBW). For CN, only the strongest hyperfine component is given.
}
\end{table}

The observations were performed with the
APEX\footnote{This publication is based on data acquired with the
Atacama Pathfinder Experiment (APEX). APEX is a collaboration between
the Max-Planck-Institut f\"ur Radioastronomie,
the European Southern Observatory,
and the Onsala Space Observatory.}
12 m telescope in Chile \citep{Gusten2006} located at the 5100 m high site on
Llano de Chajnantor.
The data were obtained during observing periods in 2005 November, 2006 April
and August. The receivers used were the facility APEX-2A \citep{Risacher2006}
and the MPIfR FLASH receivers \citep{Heyminck2006}. Typical system noise
temperatures were about 200 K -- 1000 K at 290 GHz and 350 GHz, 1000 K at 460
GHz and 5000 K at 810 GHz, respectively. The spectrometers for the observations
were Fast Fourier Transform Spectrometers (FFTS) with 1 GHz bandwidth and the
channel width for the 290--350 GHz observations was approximately 122.07
kHz (8192 channels), and for the 460 GHz and 810 GHz observations 488.28 kHz
(2048 channels).  For the observations, a position-switching mode was used with
the reference position typically 180$^{\prime\prime}$ off-source. The antenna
was focused on the available planets. IK Tau itself was strong enough to serve
as a line pointing source, thus small cross scans in the $^{12}$CO(3-2) line were done to monitor
the pointing during the observations. The telescope beam sizes (HPBW) at
frequencies of the observed molecular lines are shown in Table~\ref{table:1}. The antenna
beam efficiencies are given in Table~\ref{table:2} of \citet{Gusten2006}.

To map the circumstellar envelope in the  $^{12}$CO(3-2) line, 30 positions distributed on a 5$\times$6
grid in right ascension and declination were observed. The grid spacing was
9$^{\prime\prime}$ (half the FWHM beam size at 345 GHz). A raster mapping
procedure was used along the parallel grid lines with an integration time of 15
s.

\begin{table}
\caption{Beam efficiencies for the different receivers.}             
\label{table:2}      
\centering                          
\begin{tabular}{c c c}        
\hline\hline                 
Receiver & Forward efficiency &Beam efficiency \\    
\hline                        
   APEX-2A 290 GHz & 0.97 & 0.80 \\      
   APEX-2A 350 GHz & 0.97 & 0.73 \\
   FLASH 460 GHz   & 0.95 & 0.60 \\
   FLASH 810 GHz   & 0.95 & 0.43 \\
\hline                                   
\end{tabular}
\end{table}

\begin{figure} 
  \centering
  \includegraphics[width=5cm, angle=0] {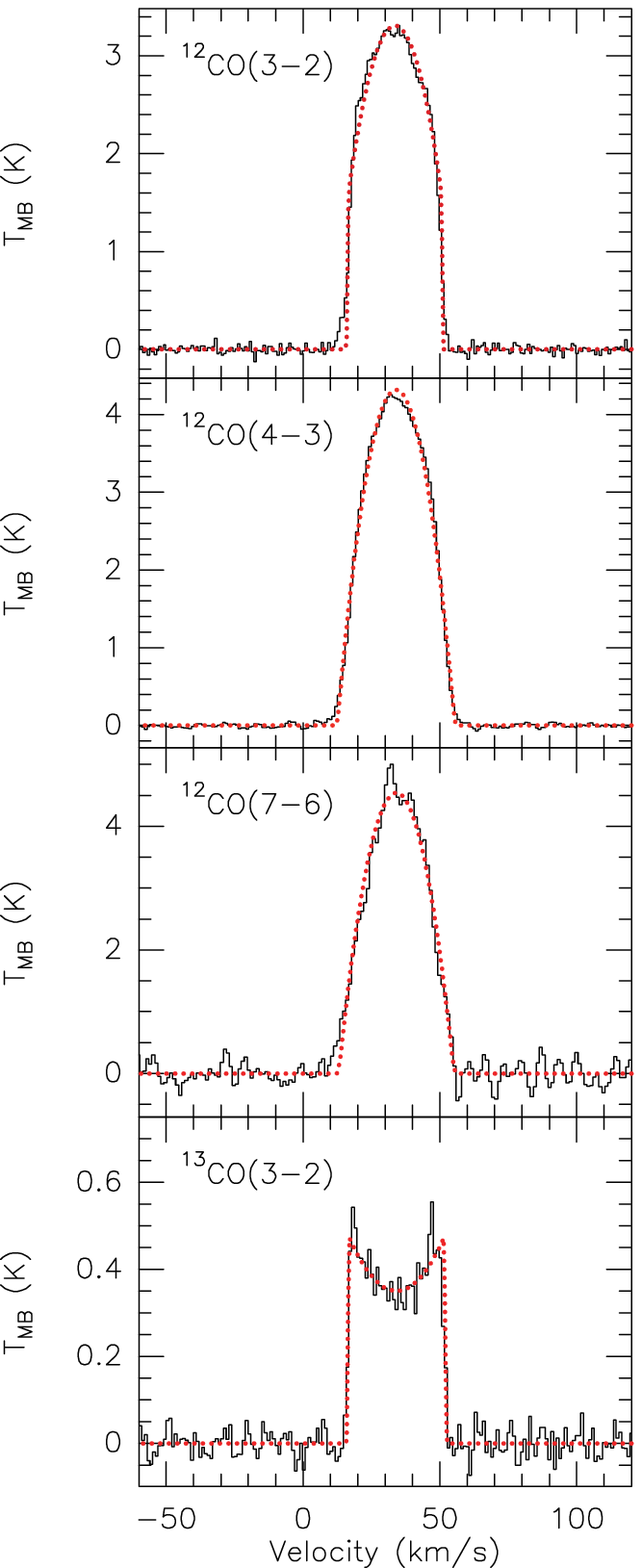}
  \caption[Observed $\mathrm{^{12}CO}$ and $\mathrm{^{13}CO}$ line profiles with expanding shell fits.]
{Observed $\mathrm{^{12}CO}$ and $\mathrm{^{13}CO}$ line profiles (solid lines) together with the expanding shell fit (dotted lines).}
  \label{co-13coplot}
\end{figure}

The spectra were reduced using the CLASS
program of the IRAM GILDAS\footnote{GILDAS is a collection of softwares oriented toward (sub-)millimeter
radio astronomical applications developed by IRAM (see more details on http://www.iram.fr/IRAMFR/GILDAS).}. To calculate the
main-beam brightness temperatures of the lines, $T_{\mathrm{MB}}$, the following
relation was used:

\begin{equation}
\label{eq3}
T_{\mathrm{MB}}={T_{\mathrm{A}}^{*}}\frac{\eta_{\mathrm{f}}}{\eta_{\mathrm{eff}}}\,.
\end{equation}

Here $T_{\mathrm{A}}^*$ is the measured antenna temperature, $\eta_{\mathrm{{f}}}$
is the forward efficiency and
$\eta_{\mathrm{{eff}}}$ is the antenna main-beam efficiency of APEX
(see Table \ref{table:2}).


\section{Observational results}

\begin{figure} 
  \centering
  \includegraphics[width=5cm, angle=0] {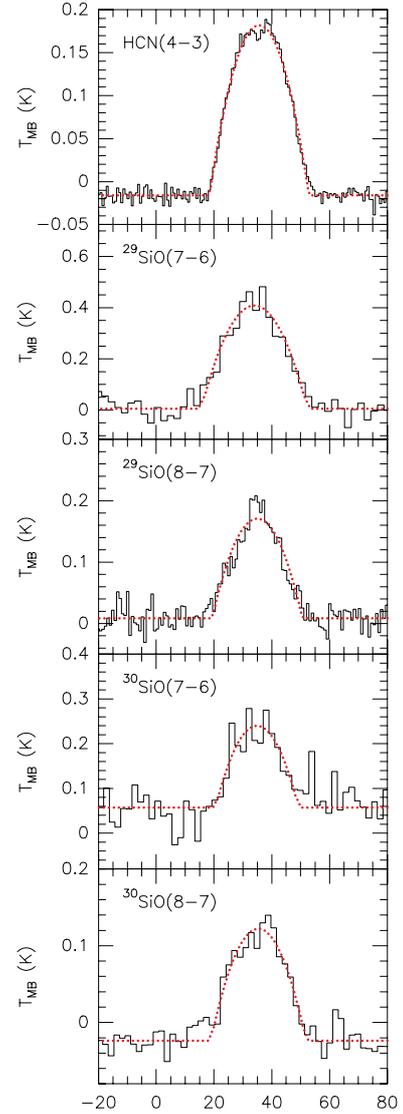}
  \caption[Observed HCN, $\mathrm{^{29}SiO}$ and $\mathrm{^{30}SiO}$
line profiles with expanding shell fits.]{Observed  HCN, $\mathrm{^{29}SiO}$ and
 $\mathrm{^{30}SiO}$ line profiles (solid lines) together with the expanding shell fit (dotted lines).}
  \label{hcn}
\end{figure}
\begin{figure*} 
  \centering
  \includegraphics[width=15cm, angle=0] {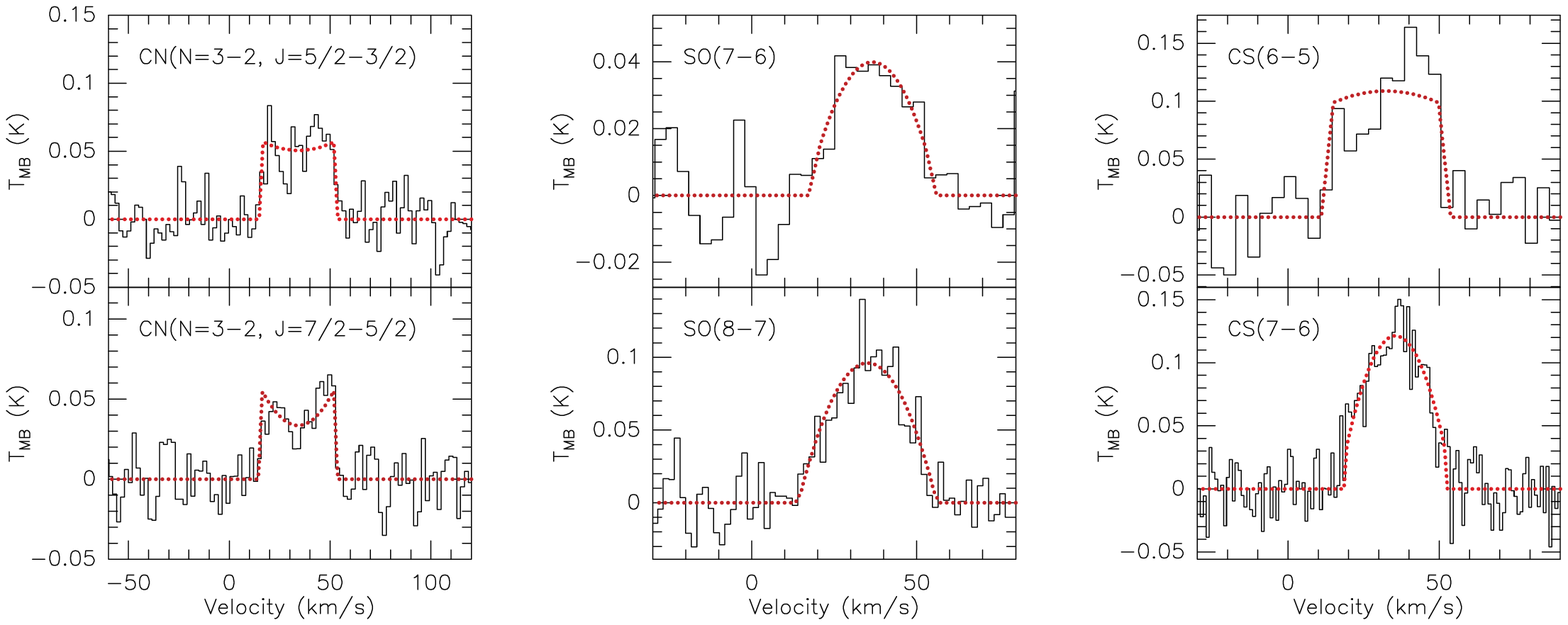}
  \caption[Observed CN, SO and CS line profiles with
  expanding shell fits.]{Observed CN, SO and CS line profiles (solid lines)
  together with the expanding shell fit  (dotted lines).}
\label{cn}
\end{figure*}
\begin{figure}
  \centering
  \includegraphics[width=5cm, angle=0] {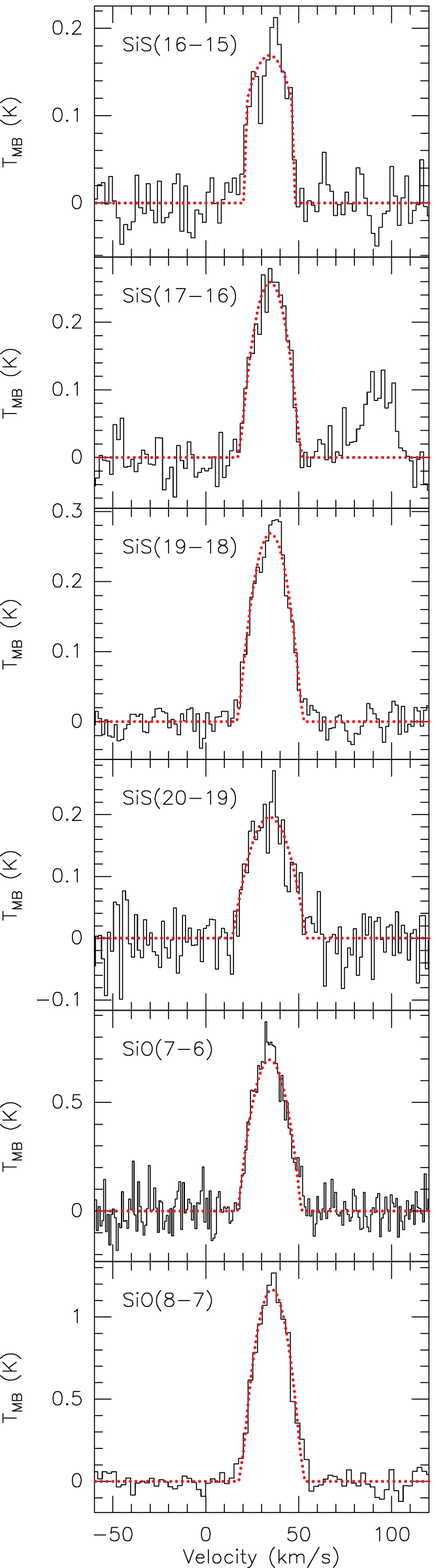}
  \caption[Observed SiS and SiO line profiles with expanding shell fit.]
{Observed SiS and SiO line profiles (solid lines) together with the expanding shell fit (dotted lines).}
  \label{sis-sioplot}
\end{figure}
\begin{figure}
  \centering
  \includegraphics[width=5cm, angle=0] {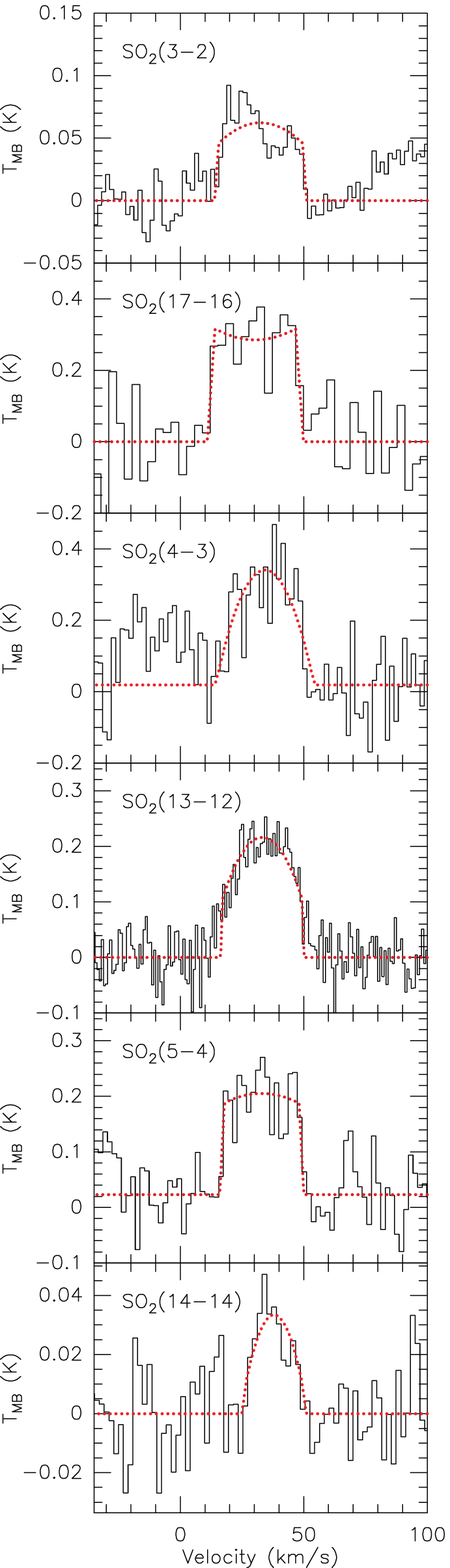}
  \caption[Observed $\mathrm{SO_{2}}$ line profiles with
  expanding shell fits.]{Observed $\mathrm{SO_{2}}$ line profiles (solid lines)
  together with the expanding shell fit  (dotted lines).}
  \label{so2}
\end{figure}

Thirty four transitions from 12 molecular species including maser lines
were detected with the APEX telescope toward IK Tau.
The detected
molecular lines are listed in Table~\ref{table:1} and their spectra are displayed in Figs.~\ref{co-13coplot}
to \ref{so2}.

\begin{figure} 
  \centering
  \includegraphics[width=5.0cm, angle=0] {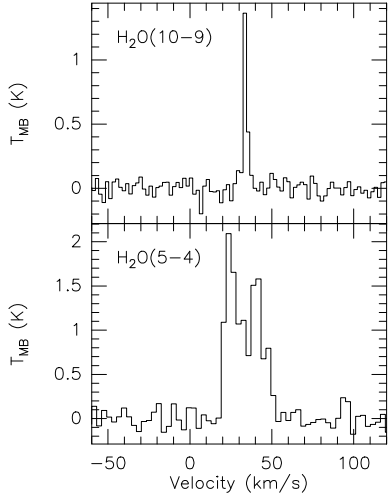}
  \caption[321 GHz and 325 GHz $\mathrm{H_{2}O}$ maser emissions observed towards IK-Tau]{321 GHz and 325 GHz
 $\mathrm{H_{2}O}$ maser emissions observed towards IK Tau.}
  \label{h2o}
\end{figure}

\begin{figure} 
  \centering
  \includegraphics[width=5.0cm, angle=0] {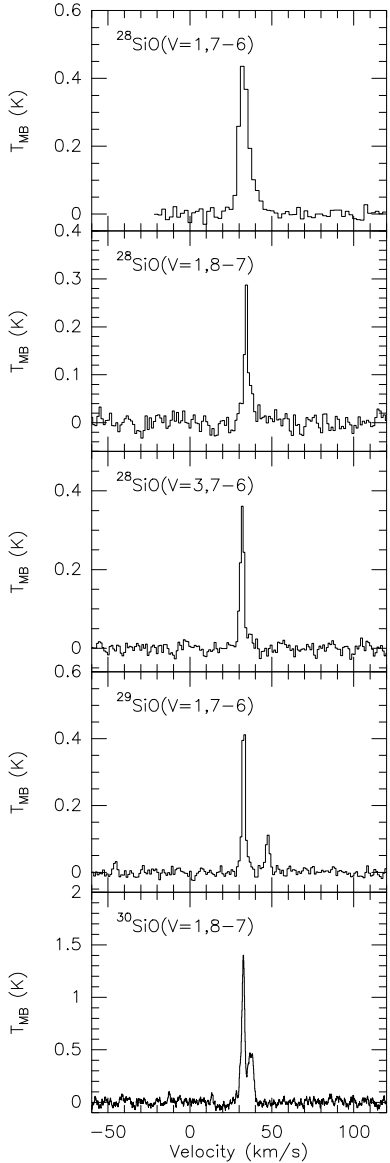}
  \caption[$\mathrm{^{28}SiO}$, $\mathrm{^{29}SiO}$ and $\mathrm{^{30}SiO}$ maser emission observed towards 
IK Tau]{$\mathrm{^{28}SiO}$, $\mathrm{^{29}SiO}$ and $\mathrm{^{30}SiO}$ maser emission detected in the rotational
 transitions $J= 7-6$ and $J=8-7$ of the fundamental $v$=1 and $v$=3 vibrational excited states observed towards IK Tau.}
  \label{sio_maser}
\end{figure}

Fig.~\ref{h2o} and Fig.~\ref{sio_maser} show the $\mathrm{H_{2}O}$ maser
lines and SiO maser lines observed toward  IK Tau, respectively; the maser line parameters are given in
 Table~\ref{maser}. 
Maser emission from $\mathrm{H_{2}O}$ at 321\,GHz and 325\,GHz was detected, as well as in the $J=7-6$ and $J=8-7$ rotational transitions within the $v=1$ and $v=3$ vibrationally excited states of $\mathrm{^{28}SiO}$, $\mathrm{^{29}SiO}$ and $\mathrm{^{30}SiO}$.

\subsection{Line parameters}


\begin{table*}
\caption{Line parameters for the detected maser lines.} 
\label{maser}      
\centering                          
\begin{tabular}{c c c c c c c c}        
\hline\hline                 
             &                               & $\mathrm{T_{MB}}$ peak (K)  & Profile $\mathrm{T_{MB}}$ area (K \, km\,s$^{-1}$) &   \\
Species   & Transition   &  Bright temp. estimate  & Bright temp. estimate  & $\mathrm{V_{lsr}}$ (km\,s$^{-1}$)    \\
\hline
H$_{2}$O     &  10$_{2\, 9}$  - 9$_{3\, 6}$ & 1.36  & 4.31   & 33.4       \\
             &  5$_{1\, 5}$  - 4$_{2\, 2}$  & 2.09  & 37.9  & 35.0       \\       
\hline
$^{28}$SiO   &  V=1, 7 - 6         & 0.44   & 3.64   & 32.5        \\
             &  V=1, 8 - 7         & 0.29   & 1.00   & 34.6       \\
             &  V=3, 7 - 6         & 0.36 & 1.22   & 33.4       \\
\hline
 $^{29}$SiO  & V=1, 7 - 6         & 0.41  & 1.63  & 32.9         \\
\hline
 $^{30}$SiO  & V=1, 8 - 7         & 1.40  & 5.4   & 33.8         \\
\hline                                   
\end{tabular}
\end{table*}
\begin{table*}
\caption{Line parameters for
each (non-masering) transition.} 
\label{lineParameter}      
\centering                          
\begin{tabular}{c c c c c c c c}        
\hline\hline                 
             &                &            &                      & $\mathrm{T_{MB}}$ peak (K) /        &
 Profile $\mathrm{T_{MB}}$ area  (K \, km\,s$^{-1}$) /&           \\
   Species   & Transition      & $\frac{\mathrm{Eu}}{\mathrm{k}}$ (K)    & $\mu_0^2$S (Debye$^2$)  & Mean $\mathrm{T_{b}}$ &
 Integrated mean $\mathrm{T_{b}}$  &  V$_{\mathrm{exp}}$ (km\,s$^{-1}$)        \\
\hline
   $^{12}$CO &   3 - 2        & 33.2  & 0.04 & 3.31 / 5.99 &  95.7 / 173  & 17.3        \\
             &   4 - 3        & 55.3  & 0.05 & 4.25 / 6.33 &  125 / 186   & 21.2      \\
             &   7 - 6        & 155   & 0.08 & 5.04 / 5.85 &  129 / 149   & 21.0       \\
\hline
   $^{13}$CO & 3 - 2  & 31.7  & 0.04 & 0.58 / 1.10 &  15.2 / 23.0 & 17.8       \\
\hline
    SiS      &  16 - 15       & 118   & 47.9 & 0.21 / 19.3 &  3.91 / 360  & 12.9       \\
             &  17 - 16       & 133   & 50.9 & 0.28 / 23.4 &  5.77 / 483  & 16.4       \\
             &  19 - 18       & 166   & 56.9 & 0.29 / 19.7 &  6.33 / 430  & 17.3        \\
             &  20 - 19       & 183   & 59.9 & 0.27 / 18.3 &   5.2 / 353  & 19.4       \\
\hline
    SiO      &  7 - 6         & 58.4  & 67.4 & 0.87 / 72.8 &  16.4 / 1373  & 17.0        \\
             &  8 - 7         & 75.0  & 77.0 & 1.27 / 86.3 &  26.7 / 1813  & 16.5       \\
\hline
    SO       & 7$_7$ - 6$_6$  & 71.0  & 16.5 & 0.06 / 5.0  &  1.09 / 91.2  & 17.7        \\
             & 8$_8$ - 7$_7$  & 87.5  & 18.9 & 0.27 / 18.3 &  4.72 / 321   & 21.0       \\
\hline
    SO$_2$   & 3$_{3\, 1}$ - 2$_{2\, 0}$    & 27.6 & 6.64 & 0.09 / 7.5  &  2.16 / 181  & 17.8   \\
             & 17$_{1\,17}$ - 16$_{0 \,16}$ & 136  & 36.5 & 0.38 / 31.8 &  11.3 / 945  & 17.9   \\
             & 4$_{3\,1}$ - 3$_{2\,2}$      & 31.3 & 6.92 & 0.07 / 5.3  &  1.41 / 107  & 19.4   \\
             & 13$_{2\,12}$ -12$_{1\,11}$   & 93.0 & 13.4 & 0.25 / 17.0 &  6.34 / 43   & 16.5  \\
             & 5$_{3\,3}$ - 4$_{2\,2}$      & 35.9 & 7.32 & 0.06 / 4.08 & 1.36 / 92.4 & 16.1   \\
             & 14$_{4\,10}$ - 14$_{3\,11}$  & 136  & 19.6 & 0.05 / 3.40 &  0.55 / 37.4 & 12.1   \\
\hline
   CS        & 6 - 5         & 49.4   & 23.1 & 0.16 / 14.7 &  4.12 / 380  & 19.1         \\
             & 7 - 6         & 65.8   & 27.0 & 0.15 / 10.2 &  3.07 / 209  & 16.6         \\
\hline
 $^{29}$SiO  & 7 - 6         & 57.6   & 67.2 & 0.22 / 18.4 & 5.08 / 425  & 18.7         \\
             & 8 - 7         & 74.1   & 76.8 & 0.27 / 18.3 &  4.72 / 321  & 15.1         \\
\hline
 $^{30}$SiO  & 7 - 6         & 56.9   & 67.2 & 0.13 / 12.0 &  2.06 / 190  & 14.1         \\
             & 8 - 7         & 73.2   & 76.8 & 0.15 / 10.2 &  2.88 / 196  & 16.4         \\
\hline
    HCN      & 4 - 3         & 42.5   & 108  & 0.69 / 15.8 &  15.3 / 249  & 17.0         \\
\hline
   CN        & N=3 - 2, J=5/2 -3/2  & 32.6  & 6.72  & 0.08 / 5.44  &  1.98 / 135 & 18.5  \\
             & N=3 - 2, J=7/2 -5/2  & 32.7  & 9.01  & 0.07 / 4.76  &  1.60 / 109 & 18.7   \\
\hline    
\end{tabular}
\tablefoot{The third column gives  the upper energy level and the
fourth column the line strength. The fifth and sixth column yield the
peak values and integrated intensity values derived from the
observed line profiles. In the last 
column, the expansion velocity is derived from the expanding shell
fit. For  the peak flux and the integrated intensity, the values
derived from the main-beam temperature ($\mathrm{T_{MB}}$) and the
mean brightness temperature estimate (Mean $\mathrm{T_{b}}$) are given.
}
\end{table*}

To get the mean brightness temperature estimates, the spectra were
corrected by the beam-filling factors assuming a CO source size of
17$^{\prime\prime}$ \citep{Bujarrabal1991},
a HCN source size of
3.85$^{\prime\prime}$
\citep{Marvel2005}
and source sizes for the
other molecules of 2.2$^{\prime\prime}$
\citep{Lucas1992}. 
Note that the CO size may be uncertain, likely underestimated, since the signal-to-noise (S/N) ratios of the profiles obtained 
by \citet{Bujarrabal1991} are much smaller than those of the CO profiles presented in this paper.
 
The beam-filling factor is given by
\begin{equation}
f=\frac{ \theta_{\mathrm{S}}^{2}}{\theta_{\mathrm{S}}^{2}+\theta_{\mathrm{b}}^{2}}\,,
\end{equation}
where $\theta_{\mathrm{S}}$ is the source size and
$\theta_{\mathrm{b}}$ is the half-power beam width (HPBW) shown in Table~\ref{table:1}. Both source and beam are assumed
to be circular Gaussians. The
mean brightness temperature estimate is computed by
\begin{equation}
T_{\mathrm{b}}=\frac{1}{f} T_\mathrm{MB}\,.
\end{equation}
Line parameters were derived with CLASS
(see more details on http://www.iram.fr/IRAMFR/GILDAS) from fitting the spectral lines with expanding shell fits, from which the expansion velocity of the envelope is obtained.
The observed maser line and thermal emission line parameters are given in Table \ref{maser} and \ref{lineParameter}, including the
envelope expansion velocity $V_{\mathrm{exp}}$, the main beam brightness temperature $T_\mathrm{MB}$, the integrated
area, and the parameters of the expanding shell fits.
The expansion velocities are distributed from 14
$\mathrm{km\,s^{-1}}$ to 21 $\mathrm{km\,s^{-1}}$.

When the S/N ratio is high enough to warrant a
consideration of the shape of the line profiles, they appear to be
characteristic for circumstellar envelopes \citep[for more detail
see][]{Zuckerman1987} : the $\mathrm{^{12}CO}$ lines have the parabolic
shape of optically thick lines and the $\mathrm{^{13}CO}$($3-2$) line has the
double-horn shape of spatially resolved optically thin lines 
(see Fig.~\ref{co-13coplot}). Lines
from the three SiO isotopologues and SiS lines have a Gaussian shape
(see Fig.~\ref{sis-sioplot}),
indicating that they are partially formed in the wind acceleration
regime where the stellar winds has not yet reached its full terminal
velocity \citep{Bujarrabal1991}. Some of the $\mathrm{SO_{2}}$ lines
 seem to show the square shape characteristic of unresolved optically thin
lines and some of them have the parabolic shape of optically thick
lines
(see Fig.~\ref{so2}). 
CS and SO lines seem to have the square shape of unresolved optically
thin lines for low excitation transitions and the parabolic shape of
optically thick lines for high excitation transitions
(see Fig.~\ref{cn}). HCN shows a
global parabolic shape with a weak double-peak profile on the top
(see Fig.~\ref{hcn}).
For the CN molecule fits to the spectra were done that take the
hyperfine structure of the molecule into account. 
 Although the S/N of the individual
components is small, the observations are not in agreement with the
optical thin ratio of different HFS components and hint to hyperfine
anomalies as already reported by \citet{Bachiller1997}.


\subsection{$\mathrm{CO}$ Maps} \label{mapCO}

\begin{figure*} 
  \centering
  \includegraphics[width=11cm, angle=0] {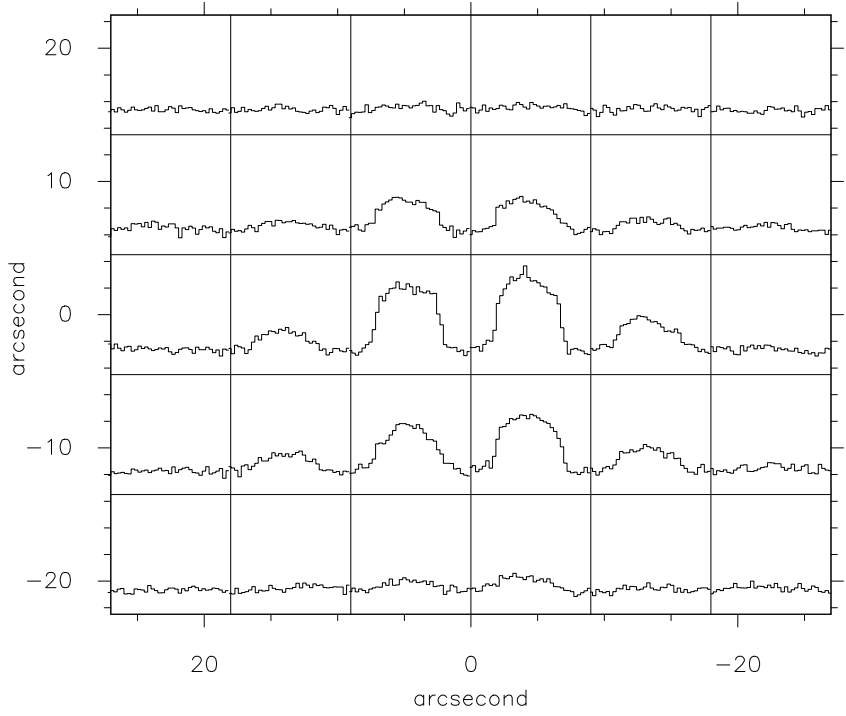}
  \caption[Map of $\mathrm{^{12}CO}$ J=3-2 line emission]{Map of
  $\mathrm{^{12}CO}$ J=3-2 line emission toward IK Tau. The grid
  spacing was 9$^{\prime\prime}$. The main-beam brightness temperatures in the spectra
  range from 0.5 K to 3.3 K.}
\vspace{2pt} 
\label{cogrid}
\end{figure*}
\begin{figure*} 
  \centering
  \includegraphics[width=11cm, angle=0] {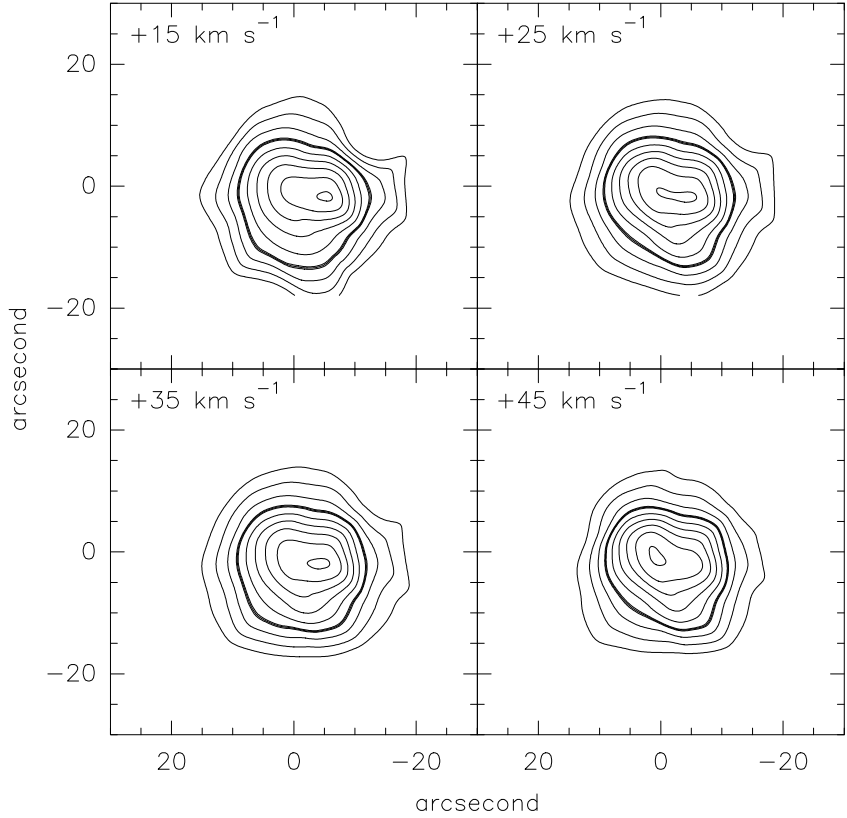}
  \caption[Contour maps of the $\mathrm{^{12}CO}$ J=3-2 line
  emission]{Contour maps of the $\mathrm{^{12}CO}$ J=3-2 line emission
    for IK Tau integrated over 10 $\mathrm{km\,s^{-1}}$ centered at
    velocities of 15 $\mathrm{km\,s^{-1}}$, 25 $\mathrm{km\,s^{-1}}$,
    35 $\mathrm{km\,s^{-1}}$ and 45 $\mathrm{km\,s^{-1}}$. One
    velocity interval is centered near the stellar radial velocity
     of 33.7 $\mathrm{km\,s^{-1}}$.The contour values are 20 $\%$ to 90
    $\%$ and 99 $\%$ of the maximum integrated intensity in each
    velocity interval, which is 7, 29, 32, and 26
    $\mathrm{K\,km\,s^{-1}}$ for the 15, 25, 35, and 45
    $\mathrm{km\,s^{-1}}$ channel, respectively.  The 50 $\%$ contour level
    is drawn in boldface.}
  \vspace{2pt}
  \label{comap}
\end{figure*}

The spectra resulting from mapping the $\mathrm{^{12}CO(3-2)}$
transition in a region of 45$^{\prime\prime}$$\times$54$^{\prime\prime}$ around IK Tau
are shown in Fig. ~\ref{cogrid}. These spectra provide us with a tool to derive
the source size as a function of radial velocity (see
Fig.~\ref{comap}).The envelope of IK Tau appears roughly spherically
symmetric in $\mathrm{^{12}CO(3-2)}$ with a deconvolved extent at
half-peak integrated intensity of 20$^{\prime\prime}$. The physical diameter of
the emission region is thus $2.1\times10^{16}$ cm assuming a source
distance of 250 pc.

\section{Modeling results}
\subsection{Physical structure of the envelope} \label{CO_modeling}


CO lines are amongst the best tools to estimate the global properties
of circumstellar envelopes, since the abundance of CO is quite
constant across the envelope, except for photo-dissociation effects at
the outer edge \citep{Mamon1988ApJ...328..797M}. The spatial
distribution of CO was found from our mapping observation to be
spherically symmetric (see Sect.~\ref{mapCO}). A detailed multi-line
non-LTE (non local thermodynamic equilibrium) study
of CO can therefore be used to determine the physical properties of
the envelope.  

The one-dimensional version of the Monte Carlo code RATRAN
\citep{Hogerheijde2000} was used to simulate the CO lines' emission.
The basic idea of the Monte Carlo method is to split the emergent
radiative energies into \emph{photon packages}, which perform a random
walk through the model volume. This allows the separation of local and
external contributions of the radiation field and makes it possible
to calculate the radiative transfer and excitation of molecular lines.
The Monte Carlo method for molecular line transfer has been described
by \citet{Bernes1979} for a spherically symmetric cloud with a uniform
density. The code is formulated from the viewpoint of cells rather
than photons.  It shows accurate and fast performance even for high
opacities \citep[for more details see][]{Hogerheijde2000}. The circumstellar envelope is assumed to be spherically
symmetric, to be produced by a constant mass-loss rate, and to expand
at a constant velocity. In the Monte Carlo simulation,
typically $10^3$ model photons are followed throughout the envelope
until they escape. The region is divided into discrete grid
shells, each with constant properties (density, temperature, molecular
abundance, turbulent line width, etc.).

For the case of a steady state, spherically symmetric outflow, the gas
density as a function of radial distance from the center of the AGB
star is given by
\begin{equation}
\label{eqDens}
n(r)=\frac{\dot{M}}{4\pi r^{2} V_{\mathrm{exp}} m}\,,
\end{equation}
where $m$ is the mass of the typical gas particle, taken to be $m
\sim 3\times 10^{-24}$ gram, since the gas is mainly in molecular form
in AGB envelopes \citep{Teyssier2006}.  

The kinetic temperature is assumed to vary as
\begin{equation}
\label{eqTemp}
T=T_{0}\left[\frac{10^{16}}{r(\mathrm{cm})}\right]^{\alpha}+T_{\mathrm{bg}}\,,
\end{equation}
where $T_{0}$ is the temperature at $1 \times 10^{16}$\,cm and
$T_{\mathrm{bg}}$ represents the background temperature. With the radial profiles for density and temperature given by Eq.~\ref{eqDens} and \ref{eqTemp}, the program
solves for the molecular excitation as a function of radius. Beside
collisional excitation, radiation from the cosmic microwave background
and thermal radiation from local dust were taken into account.
Then, the molecular emission is integrated in radial direction over
the line of sight and convolved with the appropriate antenna beam.

\begin{figure} 
  \centering
  \includegraphics[width=7.0cm, angle=0] {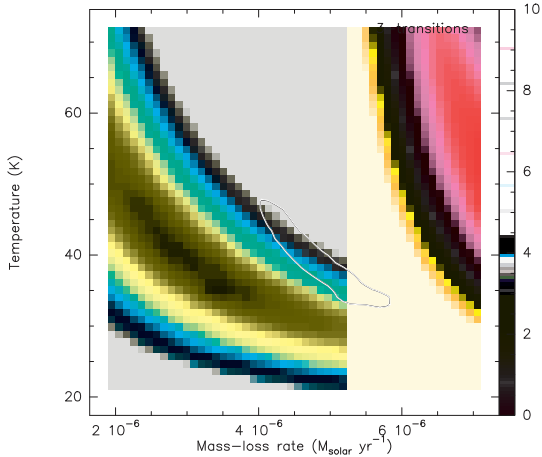}
  \caption[Reduced $\chi^2$ contour maps]{Reduced $\chi^2$ contour maps of the model to input
variable parameters, i.e., the mass-loss rate ($\dot{M}$) and 
temperature ($T_0$), which is the kinetic temperature at a given radius ($1 \times 10^{16}$\,cm). 
The contour level is drawn at 0.94
indicating the 68 $\%$ confidence level.  
  }
  \label{fitchi}
\end{figure}

The best-fit model is found by minimizing the total $\chi^2$ using
the $\chi^2$ statistic defined as
\begin{equation}
\label{eqChi}
\chi^2=\sum_{i=1}^{N}{\frac{[I_{\mathrm{mod}}-I_{\mathrm{obs}}]^2}{\sigma^2}}.
\end{equation}
where $I_{\mathrm{mod}}$ is the line intensity of the model and $I_{\mathrm{obs}}$ is
the observation, $\sigma$ is $rms$ noise of the observed spectra, the summation is done
over all channels $N$ of the three $^{12}$CO line transitions as observed for this project with APEX, i.e. $J=3-2$, $J=4-3$, and $J=7-6$.
We have put more weight on the reproduction of the
line shapes and the fitting of the lines observed
with the APEX telescope which were calibrated in a
consistent way than on the reproduction of the lines taken from the literature.
The reduced $\chi^{2}$ for the models is given by
\begin{equation}
\label{eqRedChi}
\chi^2_{\mathrm{red}}=\frac{\chi^{2}}{d.f},
\end{equation}
where $d.f$ is the degree of freedom being $N-p$, with $p$ the
number of adjustable parameters.  Fig.~\ref{fitchi} shows the
$\chi^{2}$ contour plot produced by varying the mass-loss rate and the
temperature $T_0$.  In this figure, the 68\,\% confidence limit, i.e.\
the 1$\sigma$ level, is indicated. In this region, the temperature
$T_0$ ranges between 34 to 47\,K and the mass-loss rate is in the
range from 4.0$\times$10$^{-6}$ to 5.7$\times$10$^{-6}$\,M$_\odot$/yr.

\begin{figure*} 
  \centering
  \includegraphics[width=13.0cm, angle=0] {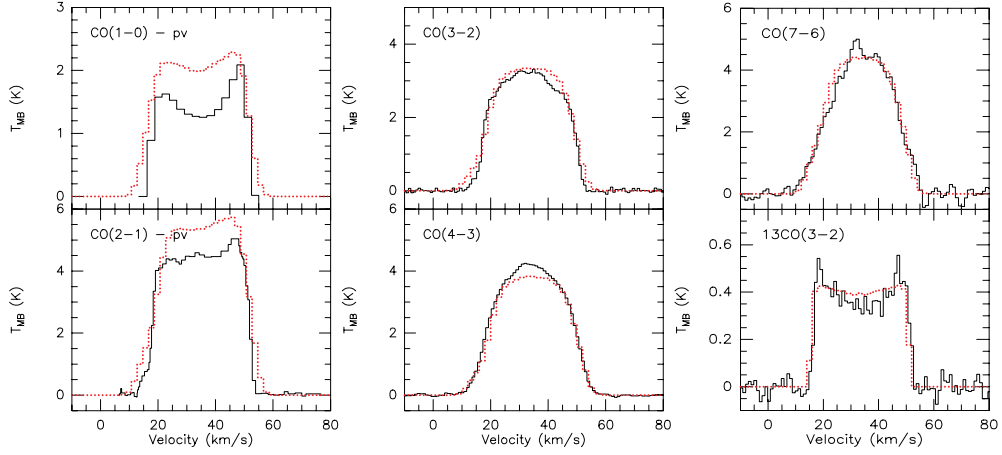}
  \caption[Best-fit model predictions for the CO transitions for IK
  Tau]{Best-fit model spectra for the different CO transitions to the available
    set of data at offset (0$^{\prime\prime}$,0$^{\prime\prime}$) for
    IK Tau. Parameters used to calculate the best-fit theoretical
    predictions are given in Table~\ref{table:7}. The
$\mathrm{^{12}CO(1-0)}$ and $\mathrm{^{12}CO(2-1)}$ data
    are from \citet{Teyssier2006}, where $^{\prime}$pv$^{\prime}$
    means the IRAM 30 m telescope in Pico Veleta. The model fits are shown by dotted lines.}
  \label{fit}
\end{figure*}

\begin{figure*}
  \centering
  \includegraphics[width=14.0cm, angle=270] {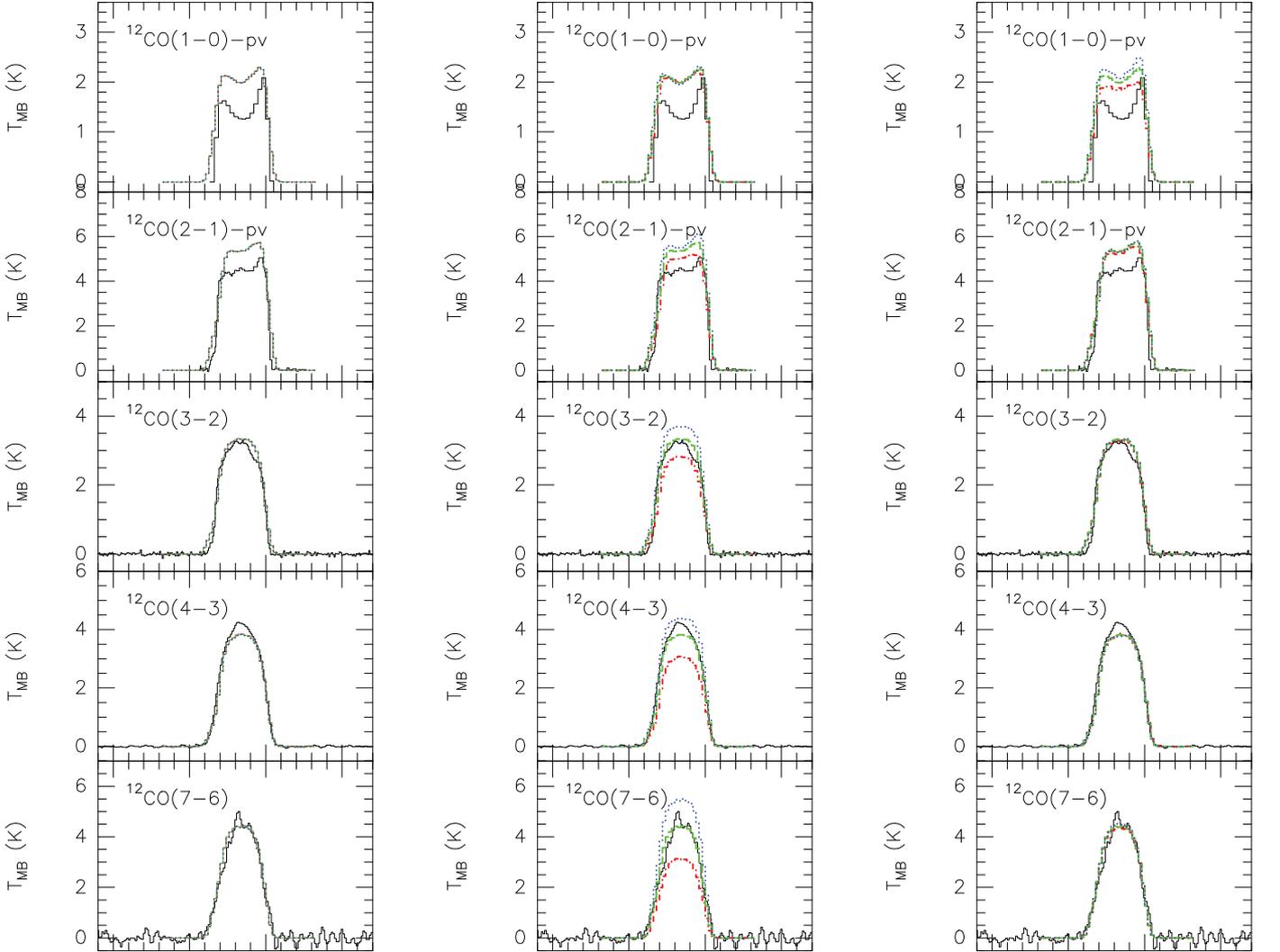}
  \caption[Model fits with different input parameters]{Model fits with
    different input parameters to the set of data at offset
    (0$^{\prime\prime}$, 0$^{\prime\prime}$) for IK Tau. Left column 
    shows variations in the inner radius:
    $0.1\times10^{14}$ cm (dash-dotted line),  $1\times10^{14}$ cm (dashed line),  $10\times10^{14}$ cm (dotted
    line). Middle column
     shows variation in $T_0$: $30$\,K (dash-dotted line),  $40$\,K (dashed line),
     $50$\,K (dotted line). Right column shows variations in the
    outer radius:  $5.3 \times10^{16}$ cm (dash-dotted line), 
    $6.3\times10^{16}$ cm (dashed line), $7.3\times10^{16}$ cm (dotted line) . }
  \label{vari}
\end{figure*}

\begin{figure*} 
  \centering
  \includegraphics[width=6.0cm, angle=270] {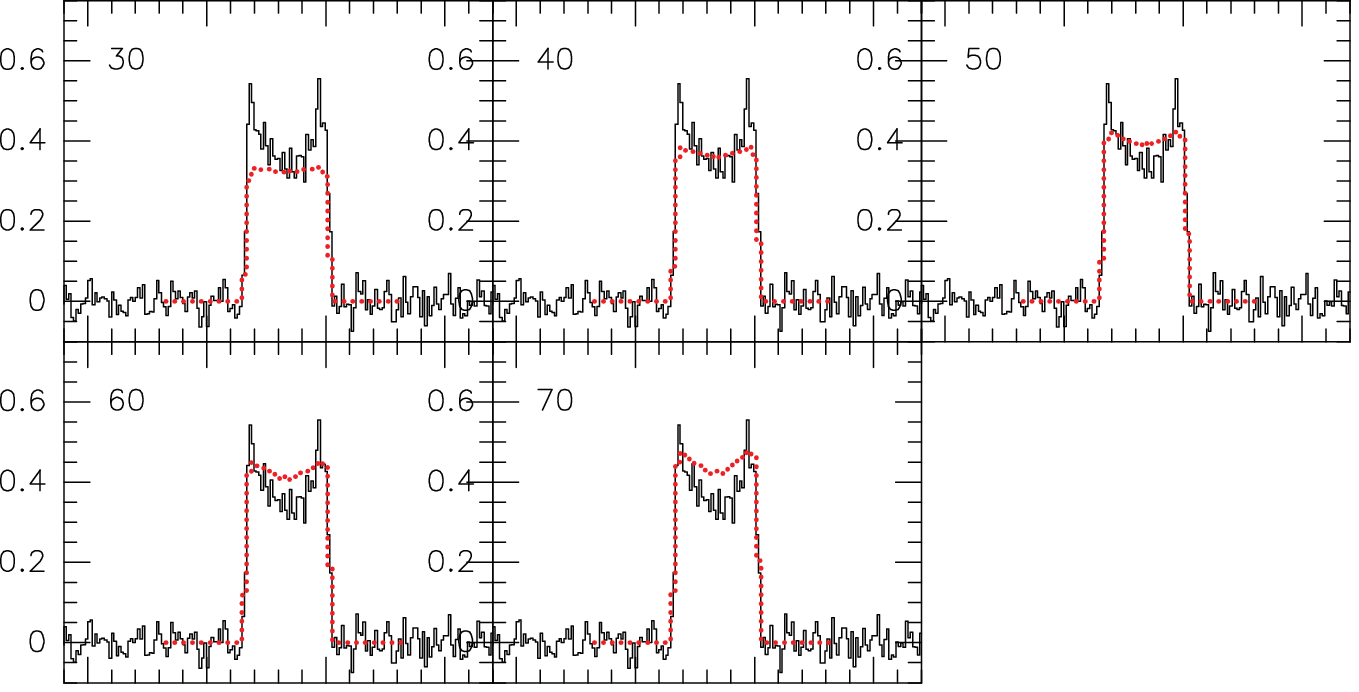}
  \caption[Model fits to the $\mathrm{^{13}CO}$(3-2) transition with
  different $T_0$] {Model fits of the $\mathrm{^{13}CO}$(3-2) transition
    with different values for $T_0$, as indicated in the upper left
    corner of each panel. The model fits are shown by dotted lines.}
  \label{fit13CO}
\end{figure*}

The best-fit model parameters are listed in Table~\ref{table:7};
the results of the model fits are shown in Fig.~\ref{fit}. In
Fig.~\ref{vari} theoretical model predictions for the
$\mathrm{^{12}CO}$ lines with different inner
radii, different $T_{0}$, and different outer radii are shown.
Predictions  for
$\mathrm{^{13}CO}$ with different $T_{0}$ are presented in Fig.~\ref{fit13CO}.
Predictions for intensities at the observed offset positions were done from the
best fit model and are consistent with the size determined from the observed
CO maps.

As shown in Fig.~\ref{fit}, the overall line profiles are fit very well
for the higher $J$ transitions ($\mathrm{^{12}CO(3-2)}$,
$\mathrm{^{12}CO(4-3)}$, $\mathrm{^{12}CO(7-6)}$). However, the model intensities of the IRAM $\mathrm{^{12}CO(1-0)}$ and $\mathrm{^{12}CO(2-1)}$ transitions  are somewhat higher than the observational data taken from literature, 
but the shapes  fit satisfactorily. The predictions for the $\mathrm{^{12}CO(2-1)}$ line are still within the absolute uncertainty of the line, but this is not the case for the $\mathrm{^{12}CO(1-0)}$ line. An obvious reason for this mismatch could be a problem with the outer radius value. However, our sensitivity analysis (see Fig.~\ref{vari} and see discussion in next paragraphs) shows that, while lowering the outer radius value 
indeed the total integrated intensity decreases, the line shape is not well reproduced anymore. Since the relative uncertainty (i.e., the line shape) is much lower than the absolute intensity (i.e., the integrated intensity), we have put more weight on the reproduction of the line shapes. Moreover,
 we note that this is not the first time that a non-compatability of the IRAM fluxes with other observed data is reported \citep[e.g.][]{Decin2008A&A...484..401D}.
The $\mathrm{^{13}CO(3-2)}$ line clearly shows a double-horn profile
and the best-fit results in a somewhat different $T_{0}$ and a
different outer radius than for the $^{12}$CO data. Nevertheless, the
best-fit value for $T_0$ derived from $\mathrm{^{12}CO}$ still gives a
reasonable fit to the $\mathrm{^{13}CO}$ line (Fig.~\ref{fit13CO}). As
shown in Fig.~\ref{fit13CO}, the intensities of the profiles do not
change so much with $T_{0}$ but the lines show a flat shape on top for
the lower temperatures (30 K and 40 K), and a double-horn shape at
higher temperatures.

As shown in Fig.~\ref{vari}, the line shapes and intensities for all
transitions are not much influenced by the inner radius variations
since the $\mathrm{^{12}CO}$ emission contributing dominantly to the
spectra arises from regions further out in the envelope. The outer
radius variations mainly affect the $\mathrm{^{12}CO(1-0)}$ line,
which is formed further out in the envelope than the other
$\mathrm{^{12}CO}$ transitions.

\begin{table*}
\centering
\caption{Parameters for the best-fit model to the observed $^{12}$CO
  and $^{13}$CO line profiles.}             
\label{table:7}      
\begin{tabular}{c l c c c c c c c c}        
\hline\hline                 
              & R$_{\mathrm{i}}$ & R$_{\mathrm{out}}$ & Mass-loss late
& f$_{\mathrm{CO}}$ & V$_{\mathrm{exp}}$ & T$_{0}$  & $\alpha$  & T$_{\mathrm{bg}}$  \\
              & (10$^{14}$ cm)   & (10$^{14}$ cm)    & (M$_{\odot}$ yr$^{-1}$)
& (10$^{-4}$) &  (km\,s$^{-1}$) & (K) &   & (K)   \\
\hline                        
$^{\mathrm{12}}$CO & 1 & 630 & 4.7$\times$10$^{-6}$ & 3 & 18 & 40 & 0.8 & 2.7 \\
$^{\mathrm{13}}$CO & 1 & 700 & 4.7$\times$10$^{-6}$ & 0.35 & 18 & 50 & 0.8 & 2.7 \\
\hline                                   
\end{tabular}
\tablefoot{${\rm{R_i}}$ represents the inner
  radius of the envelope, R$_{\mathrm{out}}$ the outer radius and
  f$_{\mathrm{CO}}$=[CO/H$_2$] the CO abundance relative to H$_2$. The
  last three columns give the values for the parameters in
  Eq.~\ref{eqTemp}. The expansion velocity is derived from the
  $^{12}$CO(3-2) observations.
}
\end{table*}

\subsection{Chemical abundance structure} \label{Abundances}


As explained in the introduction, the density distribution of each
molecule is different, depending on the chemical processes partaking
in the envelope. The \emph{fractional abundance} of a species A is
usually specified as 
\begin{equation}
f_{\mathrm{A}}(r) =\frac{n_{\mathrm{A}}(r)} {n_{\mathrm{H_{2}}}(r)}\,,
\end{equation}
where $n_{\mathrm{H_{2}}}(r)$ is the number density of
$\mathrm{H_{2}}$ and $n_{\mathrm{A}}(r)$ is the number density of 
species A.

A first order assessment of the molecular abundance fractions can be obtained assuming that the envelope structure is in local thermodynamic equilibrium. Assuming a spherically symmetric envelope, the fractional abundance for an optically thin rotational line ($J \rightarrow J-1$) of a linear rotor is given by \citet{Olofsson1991}.\\

\begin{eqnarray}
\label{eq5}
f_x&=&3.8\times10^{-16}T_{\mathrm{MB}}\frac{V^{2}_{\mathrm{exp}}BD}{\dot{M}}\frac{T_{\mathrm{ex}}}{\mu^{2}_{0}B^{2}_{0}J^{2}}
\nonumber \\ \nonumber \\
& &\cdot \exp(\frac{h(B_{0}\times10^{9})J(J+1)}{kT_{\mathrm{ex}}})
 \frac{1}{\int^{x_{e}}_{x_{i}}\exp(-4\ln{2}x^{2})dx}
\end{eqnarray}

where $T_{\mathrm{MB}}$ is the main-beam brightness temperature,
$T_{\mathrm{ex}}$ is the excitation temperature (= $T_{\mathrm{rot}}$
and equal to the kinetic temperature under the LTE assumption),
$\mu_{0}$ is the dipole moment in Debye, $B_{0}$ is the rotational
constant in GHz, $V_{\mathrm{exp}}$ is the gas expansion velocity of
the CSE in $\mathrm{km\,s^{-1}}$, $B$ is the beam size in arcseconds,
$D$ is the distance to the source in pc, $\dot{M}$ is the mass-loss
rate in $\mathrm{M_{\odot}\,yr^{-1}}$,
and $x_i$ (=0) and $x_e$ are the inner and outer radius of the CSE, respectively, measured in units of $B$.
It has for simplicity been assumed that $f_x$ is constant from $x_i$ to $x_e$ and zero elsewhere. For CN,
the relative strenghts of the different hyperfine components where taken into account.
If the line is optically thick, the value of $f_x$ estimated by the above formula is only a lower limit.

The $\mathrm{SO_2}$ abundance with respect to $\mathrm{H_2}$ is estimated using the equation given by \citet{Morris1987}:

\begin{eqnarray}
\label{eq6}
f_{\mathrm{SO_2}}&=&\frac{T_{\mathrm{a}}(\mathrm{SO_2})}{2\times10^{13}}
\frac{V_{\mathrm{exp}}^{2}BDQ(T_{\mathrm{ex}})}{\dot{M}\mu_{0}^{2}S\nu}
\nonumber \\ \nonumber \\
& & \cdot\frac{1}{\exp[-1.44E_{\mathrm{u}}/T_{\mathrm{ex}}]}\frac{1}{\int^{x_{e}}_{x_{i}}\exp(-4\ln{2}x^{2})dx}
\end{eqnarray}

where $Q(T_{\mathrm{ex}})$ is the molecular partition function
($\simeq$ $1.15T_{\mathrm{ex}}^{3/2}$, for more detail see
\citet{Omont1993}),
$E_{\mathrm{u}}$ is the energy of the upper state
of the transition, $S$ is the line strength, and $\nu$ is the
frequency of the transition.

A mass loss rate of $\dot{M}=4.7\times10^{-6}$
$\mathrm{{M_{\odot}\,yr^{-1}}}$
\citep[see Sect.~\ref{CO_modeling}, and][]{Teyssier2006}
was adopted to calculate the abundances. 
Since the outer radius of the molecular emitting region can be quite uncertain for molecules for which no observational maps exist, two different outer radii will be used for these molecules (`case A' and `case B').
For SiO, the value for the outer radius was taken to be  $2\times10^{15}$ cm (case A) and
$5\times10^{15}$ cm (case B), for the other molecules
$1\times10^{16}$ cm (case A) and $5\times10^{16}$ cm (case B) was
assumed \citep{Bujarrabal1994}. For all lines from this work,  
we adopted expansion velocities from Table \ref{lineParameter}. For lines taken from the literature (see Table~\ref{abundance}),
an expanding velocity of 18 $\mathrm{km\,s^{-1}}$ is used consistent
 with our non-LTE CO modeling of the envelope.
For the
excitation temperatures, $T_{\mathrm{ex}}$, rotational temperatures
as computed from Boltzmann diagrams are taken (see Table \ref{rotational}). Values for the upper energy level and line strength ($\mu_0^2 S$) can be found in Table~\ref{lineParameter}.

\subsubsection{Results}

Using the method outlined above, the fractional abundances of all
molecules (except CO) were determined (see Table \ref{abundance}) .

\begin{table}
\caption{Rotational temperature and beam averaged column density.
} 
\label{rotational}      
\centering                          
\begin{tabular}{l c c}        
\hline\hline                 
   Species   & T$_{\mathrm{rot}}$ (K)  & N (cm$^{-2}$)   \\      
\hline
    SiS      & 85.8 (11.1)    & 4.46$\times$10$^{15}$ (1$\times$10$^{15}$)    \\
\hline
    SiO      & 17.1 (1.0)     & 8.24$\times$10$^{15}$ (1$\times$10$^{15}$)   \\
\hline
    SO       & 27.2 (2.7)     & 6.35$\times$10$^{15}$ (2$\times$10$^{15}$)   \\
\hline
    SO$_2$   & 67.5 (6.8)     & 2.02$\times$10$^{16}$ (4$\times$10$^{15}$)  \\
\hline
 $^{30}$SiO  & 68.6 (82.3)    & 2.48$\times$10$^{14}$ (4$\times$10$^{14}$)   \\
\hline
 $^{29}$SiO  & 30.0 (15.5)    & 7.12$\times$10$^{14}$ (9$\times$10$^{14}$)     \\
\hline
   CS        & 33.9 (4.7)     & 8.89$\times$10$^{14}$ (2$\times$10$^{14}$)   \\
\hline
   HCN       & 8.3 (0.5)      & 2.27$\times$10$^{15}$ (5$\times$10$^{14}$)     \\
\hline                                   
\end{tabular}
\tablefoot{The temperature and the density were determined from the rotational diagram analysis. The uncertainties
 are given within parenthesis (no systematic errors included).}
\end{table}

\begin{table*}
\caption{Derived molecular fractional abundance for each transition.
}
\label{abundance}      
\centering                          
\begin{tabular}{c c c c c c c }        
\hline\hline                 
   Species & Transition           & Abundance  & Outer radius   &  Abundance  & Outer radius  &     Reference   \\
                &                           &  (case A) &  (case A) & (case B) & (case B) & \\
\hline
    SiS     &  (5-4)   & 1.5$\times$10$^{-6}$ & 1$\times$10$^{16} \mathrm{cm}$   & 3.8$\times$10$^{-7}$  & 5$\times$10$^{16} \mathrm{cm}$        &     (1)   \\
            &  (16-15) & 6.0$\times$10$^{-7}$  &   & 1.7$\times$10$^{-7}$  &      &     \\
            &  (17-16) & 1.2$\times$10$^{-6}$  &   & 3.6$\times$10$^{-7}$  &      &     \\
            &  (19-18) & 1.3$\times$10$^{-6}$  &   & 4.2$\times$10$^{-7}$  &      &     \\
            &  (20-19) & 1.7$\times$10$^{-6}$  &   & 5.4$\times$10$^{-7}$  &      &     \\
\hline
    SiO     &  (2-1)   & 1.7$\times$10$^{-5}$  & 2$\times$10$^{15} \mathrm{cm}$    &   6.7$\times$10$^{-6}$  & 5$\times$10$^{15} \mathrm{cm}$ &       (1)  \\
            &  (3-2)   & 1.5$\times$10$^{-5}$  &   &  6.2$\times$10$^{-6}$    &    &    (1)   \\
            &  (5-4)   & 3.8$\times$10$^{-6}$  &   &  1.5$\times$10$^{-6}$    &    &    (2)   \\
            &  (7-6)   & 8.3$\times$10$^{-6}$  &   &  3.3$\times$10$^{-6}$    &    &         \\
            &  (8-7)   & 1.9$\times$10$^{-5}$  &   &  7.6$\times$10$^{-6}$    &    &         \\
\hline
    SO &  (2$_2$-1$_1$) & 2.0$\times$10$^{-7}$ & 1$\times$10$^{16} \mathrm{cm}$    &  5.2$\times$10$^{-8}$  & 5$\times$10$^{16} \mathrm{cm}$ & (2)   \\
       &  (5$_6$-4$_5$) & 1.1$\times$10$^{-6}$ &    &  4.3$\times$10$^{-7}$  &   &  (1)   \\
       &  (7$_7$-6$_6$) & 2.3$\times$10$^{-7}$ &    &  6.6$\times$10$^{-8}$  &    &       \\
       &  (8$_8$-7$_7$) & 1.6$\times$10$^{-6}$ &    &  5.1$\times$10$^{-7}$   &    &  \\
\hline
    SO$_2$  & (3$_{1\,3}$-2$_{0\,2}$)     & 1.7$\times$10$^{-5}$ & 1$\times$10$^{16} \mathrm{cm}$     & 4.3$\times$10$^{-6}$ & 5$\times$10$^{16} \mathrm{cm}$ &       (2)   \\
            & (10$_{1\,9}$-10$_{0\,10}$)  & 1.1$\times$10$^{-5}$  &   &   2.9$\times$10$^{-6}$  &   &    (2)   \\
            & (10$_{0\,10}$-9$_{1\,9}$)   & 1.6$\times$10$^{-5}$  &   &   5.1$\times$10$^{-6}$  &   &    (2)   \\
            & (3$_{3\,1}$-2$_{2\,0}$)     & 6.0$\times$10$^{-6}$  &   &   1.7$\times$10$^{-6}$  &   &         \\
            & (17$_{1\,17}$-16$_{0\,16}$) & 4.7$\times$10$^{-5}$  &   &   1.4$\times$10$^{-5}$  &   &      \\
            & (4$_{3\,1}$-3$_{2\,2}$)     & 4.8$\times$10$^{-6}$  &   &   1.5$\times$10$^{-6}$  &   &       \\
            & (13$_{2\,12}$-12$_{1\,11}$) & 2.1$\times$10$^{-5}$  &   &   6.5$\times$10$^{-6}$  &   &       \\
            & (5$_{3\,3}$-4$_{2\,2}$)     & 2.5$\times$10$^{-6}$  &   &   8.0$\times$10$^{-7}$  &   &       \\
            & (14$_{4\,10}$-14$_{3\,11}$) & 3.8$\times$10$^{-6}$  &   &   1.2$\times$10$^{-6}$  &   &       \\
\hline
 $^{30}$SiO & (7-6)    &  2.2$\times$10$^{-6}$  &  2$\times$10$^{15} \mathrm{cm}$    & 8.7$\times$10$^{-7}$    &  5$\times$10$^{15} \mathrm{cm}$   &      \\
            & (8-7)    &  6.7$\times$10$^{-6}$  &   & 2.8$\times$10$^{-6}$    &       &     \\
\hline
 $^{29}$SiO & (7-6)    &  6.2$\times$10$^{-6}$  &  2$\times$10$^{15} \mathrm{cm}$   &  2.5$\times$10$^{-6}$    & 5$\times$10$^{15} \mathrm{cm}$    &     \\
            & (8-7)    &  1.1$\times$10$^{-5}$  &   &  4.6$\times$10$^{-6}$    &     &      \\
\hline
   CS       & (2-1)   & 4.7$\times$10$^{-7}$  & 1$\times$10$^{16} \mathrm{cm}$    &   1.1$\times$10$^{-7}$  & 5$\times$10$^{16} \mathrm{cm}$    &    (3)   \\
            & (3-2)   & 1.9$\times$10$^{-7}$  &  & 5.9$\times$10$^{-8}$    &    &     (1)   \\
            & (6-5)   & 3.2$\times$10$^{-7}$  &  & 9.2$\times$10$^{-8}$    &    &          \\
            & (7-6)   & 2.0$\times$10$^{-7}$  &  & 6.3$\times$10$^{-8}$    &    &          \\
\hline
   HCN      & (1-0)   & 4.9$\times$10$^{-7}$ & 1$\times$10$^{16} \mathrm{cm}$   &   1.3$\times$10$^{-7}$  & 5$\times$10$^{16} \mathrm{cm}$      &          (1)   \\
            & (4-3)   & 2.3$\times$10$^{-6}$  &  &  7.2$\times$10$^{-7}$    &     &        \\
\hline
   CN       & N=3-2, J=5/2-3/2   & 9.8$\times$10$^{-8}$ & 1$\times$10$^{16} \mathrm{cm}$   &   3.1$\times$10$^{-8}$  & 5$\times$10$^{16} \mathrm{cm}$      &             \\
            & N=3-2, J=7/2-5/2   & 2.3$\times$10$^{-7}$  &  &  7.1$\times$10$^{-8}$    &     &        \\
\hline                                   
\end{tabular}
\tablefoot{For some molecules, other line transition were searched in literature.}
\tablebib{
(1)~\citet{Bujarrabal1994}; (2)  \cite{Omont1993}; (3) \citet{Lindqvist1988}.
}
\end{table*}

The most uncertain parameters used to derive the fractional abundances
are $T_{\mathrm{ex}}$, $D$ and $x_{e}$ (the outer
radius). $T_{\mathrm{ex}}$ is obtained from the rotational diagram
analysis, $D$ is taken from the literature, and the outer radius of
$x_{e}$ has been adopted differently for each individual molecule. We
also note that our analysis assumes optically thin emission, which is
not always the case for the studied line profiles. The line opacity
is expected to be larger for higher $J$ rotational transitions, so
that lower $J$ rotational transitions are expected to better probe the
fractional abundance.

\section{Discussion}

\begin{table*}
\caption{Comparison of the derived molecular fractional abundances with other published results.}
\label{abun-com}      
\centering                          
\begin{tabular}{c c c c c c c c c c}        
\hline\hline                 
          &  CS & HCN & SiO & SiS  & SO & SO$_2$  & CN \\
\hline
This work (case A) &  3.0$\times$10$^{-7}$ &
1.4$\times$10$^{-6}$ & 1.3$\times$10$^{-5}$ & 1.3$\times$10$^{-6}$ &
7.8$\times$10$^{-7}$ &  1.4$\times$10$^{-5}$     & 1.6$\times$10$^{-7}$    \\
This work (case B) & 8.1$\times$10$^{-8}$ &
4.3$\times$10$^{-7}$ & 5.1$\times$10$^{-6}$ & 3.7$\times$10$^{-7}$ & 2.7$\times$10$^{-7}$ &  4.2$\times$10$^{-6}$    & 5.1$\times$10$^{-8}$      \\
(1) &  1.0$\times$10$^{-7}$ & 9.8$\times$10$^{-7}$ & 1.7$\times$10$^{-5}$ & 4.4$\times$10$^{-7}$ & 2.6$\times$10$^{-6}$ &  -     & -       \\
(2) &  3.0$\times$10$^{-7}$ & 6.0$\times$10$^{-7}$ & -  & 7.0$\times$10$^{-7}$ & - &  -      & -    \\
(3) &  - & -  & 3.0$\times$10$^{-6}$ & - & 1.8$\times$10$^{-6}$ &  4.1$\times$10$^{-6}$      & -   \\
\hline
(4) &   2.9$\times$10$^{-7}$  &  1.4$\times$10$^{-7}$   & 3.2$\times$10$^{-5}$ &  3.5$\times$10$^{-6}$  & 9.1$\times$10$^{-7}$ &  2.2$\times$10$^{-7}$ & $3\times10^{-7}$    \\
(5) &   2.8$\times$10$^{-7}$  &  2.1$\times$10$^{-6}$   & 3.8$\times$10$^{-5}$ &  3.8$\times$10$^{-10}$  & 7.8$\times$10$^{-8}$ &  -    & $2.4\times10^{-10}$      \\
\hline                                   
\end{tabular}
\tablefoot{In the first part, we list results derived from an observational analysis, in the second part theoretical predictions from chemical models are given. Abundances from the chemical models by \citet{Duari1999} were selected
at a radius of $2.2R_{*}$ ($5\times10^{13}$ cm).}
\tablebib{
(1)~\citet{Bujarrabal1994}; (2) \citet{Lindqvist1988}; (3) \cite{Omont1993}; (4) \citet{Willacy1997}; (5) \citet{Duari1999}.
}
\end{table*}


\begin{figure*}
  \centering
  \includegraphics[width=\textwidth, angle=0] {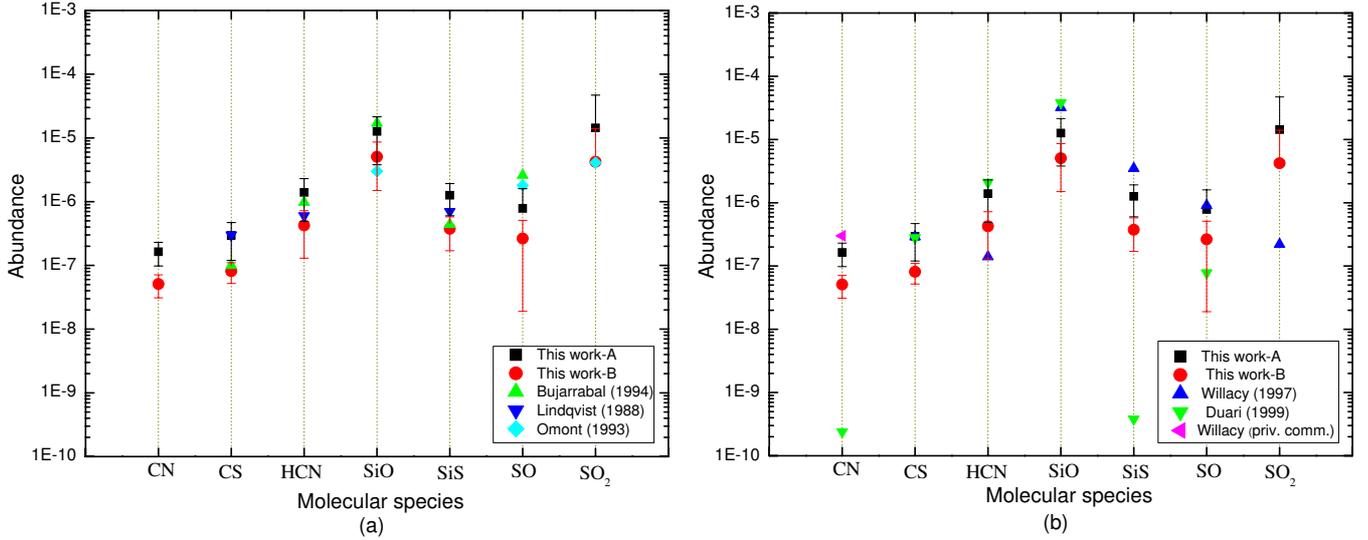}
  \caption[Abundance comparison of this work (A and B) to the
  literature]{Comparison of the molecular fractional abundances derived in this work (case A and case B-study) and values found in  literature
  \citep{Bujarrabal1994,Lindqvist1988,Omont1993, Willacy1997, Duari1999}.
  The errors are estimated from the abundance
  variations for the different transitions (see Table \ref{abun-com}). Panel
  (a) gives a comparison to other observational results, panel (b) to chemical model predictions.}
  \label{comparison}
\end{figure*}

Table \ref{abun-com} and Fig.~\ref{comparison} compare the average abundance of each molecule to values found in literature. Compared to observational results from literature \citep{Bujarrabal1994,Lindqvist1988,Omont1993},
our deduced fractional abundances agree within a factor of 3.5 for the smaller outer radius (case A) and for the larger outer radius (case B) within a factor of 10.
Compared to the predicted abundances from theoretical chemical models
by \citet{Willacy1997} and \citet{Duari1999},
we found that the predictions are comparable to our deduced values
(using the smaller outer radius, case A) within a factor of $\sim$3  for 
$\mathrm{CS}$  
and $\mathrm{SiO}$. 
Our deduced value for the SO, SiO, CN, and SiS fractional abundances agree with the results of \citet{Willacy1997}, but the predicted values by \citet{Duari1999} are much lower. The $\mathrm{SO_2}$ abundance from this work
is almost two orders of magnitude higher than the value predicted by \citet{Willacy1997}.

 As noted above, the SiS abundance in the chemical models of \citet{Duari1999} is much lower than the observed value. The chemical models by \citet{Duari1999} focus on the inner envelope (within few stellar radii), while \citet{Willacy1997} studied the chemical processes partaking in the outer envelope. The agreement between our deduced value for the fractional abundance of SiS and the predictions by \citet{Willacy1997} suggests that SiS is formed in the outer envelope.

The deduced SO abundance is a factor $\sim$10 higher than the inner wind predictions by \citet{Duari1999}, but agree with the outer wind predictions by \citet{Willacy1997}. \citet{Willacy1997} assumed no SO injection, but only in-situ formation. CN is clearly produced in the outer envelope, as photo-dissociation product of HCN.

The abundance of $\mathrm{SO_2}$ found by \citet{Willacy1997} is
much lower than the observed ones. A value of $1.4 \times 10^{-5}$ (case A) means that SO$_2$ contains 80\,\% of the solar sulphur value.
\citet{Willacy1997} suggest that SO$_2$ may be formed in a different part of the envelope compared to the other sulphur bearing molecules, for example in shocks in bipolar outflow or in the inner envelope. An indication for the typical behavior of 
$\mathrm{SO_2}$ comes also from the line profiles, e.g. the $\mathrm{SO_2}$ (14--14) line is clearly narrower and
shifted to the red.

The SiO abundance derived in this study is close to the abundance predicted
by the theoretical chemical models.
\citet{Cherchneff2006} investigated the non-equilibrium chemistry of the inner
winds of AGB stars and derived an almost constant, high SiO abundance (about
$4\times10^{-5}$ before the condensation of dust). \citet{Duari1999} and \citet{Willacy1997} derived
$3.8\times10^{-5}$ and $3.2\times10^{-5}$ for the inner and outer wind, respectively. 
Furthermore, \citet{Gonzalez2003} performed an extensive radiative
transfer analysis of circumstellar SiO emission from a large sample of
M-type AGB stars, where they adopted the assumption that the gas-phase
SiO abundance stays high close to the star, and further out the SiO
molecular abundance fraction decreases due to absorption onto dust
grains. Their results show that the derived abundances are always
below the abundances expected from stellar atmosphere equilibrium
chemistry.  For a mass-loss rate of $4 \times 10^{-6}$\,M$_\odot$/yr,
the equilibrium chemistry abundance of SiO is $\sim$3.5$\times10^{-5}$
\citep{Cherchneff2006}. Taking the scenario of depletion due to dust
formation into account, the higher excitation SiO(8-7) would probe a
higher SiO abundance. As seen in Table \ref{abundance}, the SiO(8-7) indeed probes a
higher fractional abundance, although not significantly higher than
the other lines.

\section{Conclusions}\label{Conclusions}

In this work, we present for the (sub)millimeter survey
for an oxygen-rich evolved AGB star, being IK Tau, in order to study 
the chemical composition in the envelope around the central target. 

An extensive non-LTE radiative transfer analysis of circumstellar CO
was performed using a model with a power law structure in temperature
and density and a constant expansion. The observed line profiles of
$\mathrm{^{12}CO(3-2)}$, $\mathrm{^{13}CO(3-2)}$,
$\mathrm{^{12}CO(4-3)}$, and $\mathrm{^{12}CO(7-6)}$ are fit very well
by our model, yielding a mass-loss rate of $4.7 \times
10^{-6}$\,M$_\odot$/yr. 
The line shapes and intensities for all
$\mathrm{^{12}CO}$ transitions are not much influenced by variations
of the inner radius, which is understandable since the bulk of the
$\mathrm{^{12}CO}$ emission is produced in the outer envelope. The
intensities for the higher excitation CO lines depend strongly on the
assumed temperature but not on the value of the outer radius.

For 7 other molecules (SiO, SiS, HCN, CS, CN, SO, and SO$_2$) a fractional
abundance study based on the assumption of LTE is performed. A full
non-LTE analysis of all molecules is out of the scope of this
observational paper, but will be presented in a next paper \citep{Decin2010}.  
This study shows that IK Tau is a good laboratory to
study the conditions in circumstellar envelopes around oxygen-rich
stars with submillimeter-wavelength molecular lines.  The improved
abundance estimates of this study will allow refinements of the
chemical models in the future.

Molecular line modeling predicts the abundance of each molecule
as a function of radial distance from the star, although some ambiguity 
about an inner or outer wind formation process often exists.
To  get a clear picture on the different chemistry processes partaking
 in the different parts in the envelope,
mapping observations for molecules other than CO should
be performed. Since most of the submillimeter emission from molecules 
less abundant than CO  probably arises from  the inner part
of the envelope at 2 -- 4$^{\prime\prime}$ 
meaningful observations require interferometers such as the future
Atacama Large Millimeter Array (ALMA).

\space
\space

\begin{acknowledgements}This publication is based on data acquired with the Atacama Pathfinder
Experiment (APEX). APEX is a collaboration between the
Max-Planck-Institut f\"ur Radioastronomie, the ESO, and the Onsala Space
Observatory. We are grateful to APEX staff for their assistance with the observations. 
LD acknowledges support from the Fund of
  Scientific Research, Flanders, Belgium.
\end{acknowledgements}

\bibliographystyle{aa}
\bibliography{bibliography2}

\end{document}